# Validity of the linear viscoelastic model for a polymer cylinder with ultrasonic hysteresis–type absorption in a non-viscous fluid

F.G. Mitri, Z.E.A. Fellah

A B S T R A C T

A necessary condition for the validity of the linear viscoelastic model for a (passive) polymeric cylinder with an ultrasonic hysteresis-type absorption submerged in a non-viscous fluid requires that the absorption efficiency is positive ($Q_{abs} > 0$), in agreement with the law of the conservation of energy. This condition imposes restrictions on the values attributed to the normalized absorption coefficients for the compressional and shear-wave wavenumbers for each partial-wave mode $n$. The "forbidden" values produce negative axial radiation force, absorption and extinction efficiencies, as well as an enhancement of the scattering efficiency, not in agreement with the conservation of energy law. Numerical results for the radiation force, extinction, absorption and scattering efficiencies are performed for three viscoelastic (VE) polymer cylinders immersed in a non-viscous host liquid (i.e. water) with particular emphasis on the shear-wave absorption coefficient of the cylinder, the dimensionless size parameter and the partial-wave mode number $n$. Mathematical and physical constraints are established for the non-dimensional absorption coefficients of the longitudinal and shear waves for a cylinder (i.e. 2D case) and a sphere (i.e. 3D case) in terms of the sound velocities in the VE material. The analysis suggests that the domain of validity for any viscoelastic model describing acoustic attenuation inside a lossy cylinder (or sphere) in a non-viscous fluid must be verified based upon the optical theorem.

# 1. Introduction

Mathematical modeling of absorption and scattering by polymer viscoelastic (VE) materials subjected to acoustical waves is an active field of research in various fields including biomedical ultrasound for drug delivery and the design of layered particles [1], elongated bone samples [2] and micro-vessel [3] characterization to name a few areas. For a mathematical model describing the behavior of an isotropic VE polymer-type material in which the stress varies linearly with strain, a hysteresis-type of absorption has been assumed and validated experimentally [4]. Examples for such materials include polymethylmethacrylate (PMMA), polyethylene (PE), and phenolic polymer (PP) [4, 5]. For such materials, absorption varies linearly with frequency, thus, the attenuation of compressional and shear-waves inside the material core can be modeled by introducing complex wavenumbers [6] into the acoustic scattering theory, such that the expressions for the complex wavenumbers corresponding to the compressional and shear-waves contain normalized absorption coefficients independent of frequency [7, 8], which account for the attenuation of acoustical waves within the material of the scattering object.

This linear VE model assuming a hysteresis-type of absorption in plastic polymers has been utilized in various investigations dealing with numerical computations for the acoustic scattering [9-12], and radiation forces on a sphere or cylinder [13-18] in the development of improved acoustical tweezers devices [19-24], that can be coated by a layer of a VE material [25-29] in a non-viscous fluid. Moreover, the acoustic radiation force and torque experienced by a VE polymer sphere or spherical shell filled with water or air in the field of a Bessel (vortex) beam has been evaluated, based on this VE model [30-32].

Although numerical results [excluding the long-wavelength (Rayleigh) limit] for the acoustic backscattering from a VE PMMA sphere immersed in a non-viscous fluid obtained throughout this linear model are in good agreement with experimental data [11], numerical predictions for the acoustic radiation force and efficiencies using normalized absorption coefficients for the compressional $\gamma_L$ and shear-waves $\gamma_S$ extracted from empirical data (for high-density PE) [33] using the same mathematical VE model have shown anomalous effects for a certain range of values and choice of parameters (discussed in the following). Inconsistencies have been also noted recently [34]. In another example, a VE-PE layer coating an elastic gold sphere [35] and having normalized compressional and shear absorption coefficients derived from ultrasonic measurements [4], induced an anomalous negative radiation force computed in the long-wavelength limit ($ka \ll 1$).

The aim of this communication is to investigate related effects from the standpoint of the acoustic radiation force and acoustical efficiencies for a VE cylinder submerged in non-viscous water. The validity of the VE model is determined via the optical theorem for acoustical beams in cylindrical coordinates [36] used here for acoustical plane waves, that is derived based upon the law of the conservation of energy [37], stating that the extinction efficiency is the sum of the absorption and scattering efficiencies. Since the physical phenomena involve the interaction of acoustical waves with a passive object, i.e. one with no active sources of energy present in its interior, the energy should decrease by virtue of the passive absorption properties (i.e. viscous losses) of viscoelastic materials (knowing as passivity [38]), leading to positive absorption efficiency (or power). Though the analysis here is focused on the case of a cylinder, it can be directly extended to the case of a sphere or other geometries.

**2. Method, numerical results and discussions**

The interaction of acoustical waves with a cylinder involves the introduction of various concepts used to explain physical phenomena, such as the generation of the radiation force [39, 40], the acoustic extinction, and scattering in the host fluid [41-43]. The analysis is started by considering the analytical expression for the radiation force function $Y_p$ of plane progressive waves incident upon a VE cylinder as [16, 44],

$$Y_p = -\frac{2}{ka} \sum_{n=0}^{\infty} \left[ \alpha_n + \alpha_{n+1} + 2\left( \alpha_n \alpha_{n+1} + \beta_n \beta_{n+1} \right) \right], \quad (1)$$

where the coefficients $\alpha_n = \text{Re}(C_n)$ and $\beta_n = \text{Im}(C_n)$ are the real and imaginary parts of the scattering coefficients $C_n$ [45] used in the description of the scattered field from a cylinder. The parameter $k$ is the wavenumber and $a$ is the cylinder radius.

Moreover, the expressions for the efficiency factors, representing the relative amount of the energy that is extinguished, absorbed and scattered by the VE cylinder are considered. They are expressed, respectively, as [36],

$$Q_{ext} = -\frac{2}{ka}\sum_{n=0}^{\infty} \varepsilon_n \alpha_n, \qquad (2)$$

$$Q_{abs} = -\frac{2}{ka}\sum_{n=0}^{\infty} \varepsilon_n \left(\alpha_n + \alpha_n^2 + \beta_n^2\right), \qquad (3)$$

$$Q_{sca} = \frac{2}{ka}\sum_{n=0}^{\infty} \varepsilon_n \left(\alpha_n^2 + \beta_n^2\right), \qquad (4)$$

where $\varepsilon_n = 2 - \delta_{n0}$, and $\delta_{ij}$ is the Kronecker delta function.

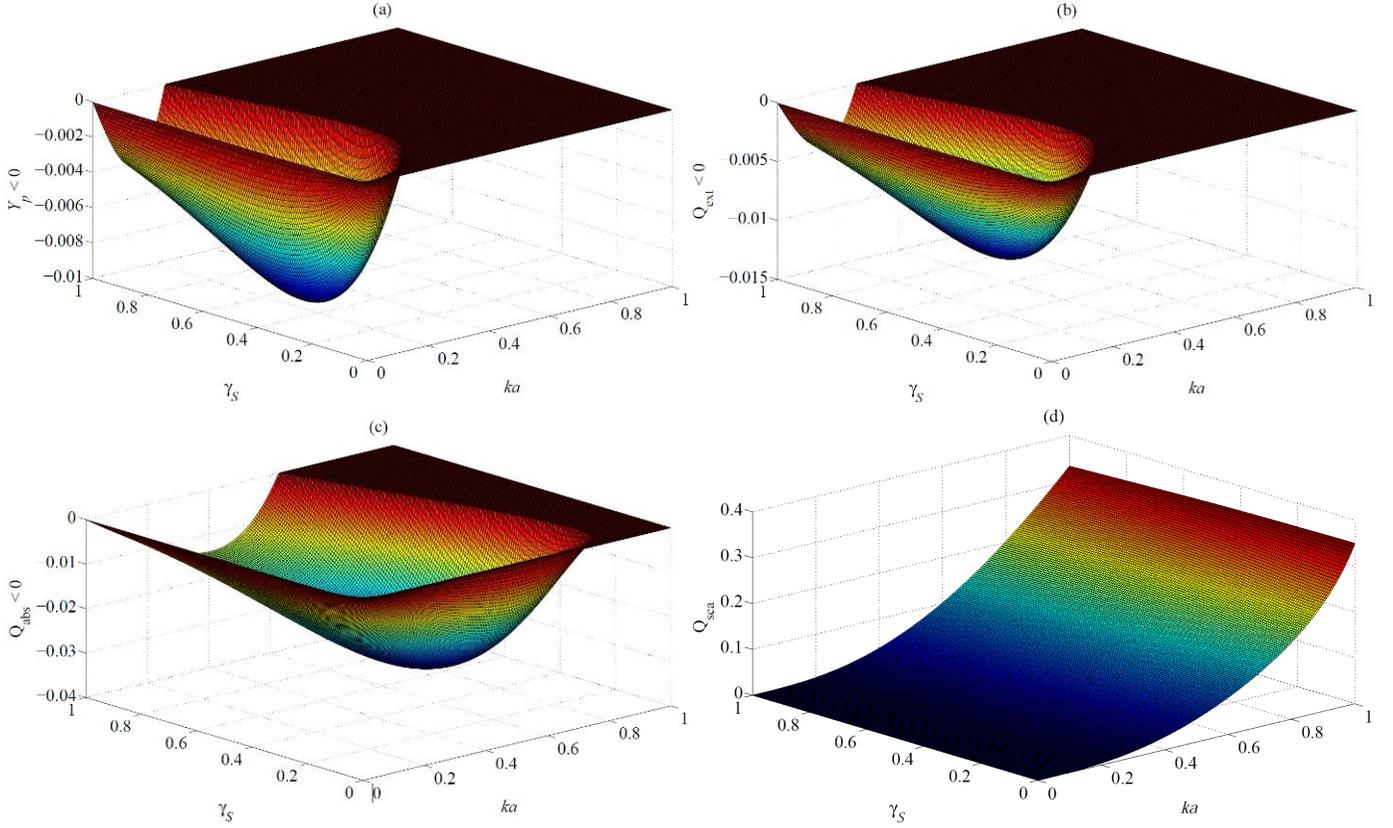

Fig. 1. Panel (a) shows the plot for the radiation force function of plane progressive waves for a PMMA VE cylinder in water. Panels (b)-(d) correspond to the plots for the extinction, absorption and scattering efficiencies, respectively. The upper limit (in amplitude) in the vertical axis has been set to zero in panels (a)-(c) for improved visualization.

The scattering coefficients of the cylinder $C_n$ are expressed in terms of cylindrical Bessel, Neumann and Hankel functions and their derivatives, and its expression is available in standard literature [42].

Consider first the case of a VE-PMMA cylindrical material (having a density $\rho_{PMMA} = 1191$ kg/m$^3$, with a speed of sound for the compressional $c_{L,PMMA} = 2690$ m/s and shear waves $c_{S,PMMA} = 1340$ m/s), immersed in (non-viscous) water. Absorption by the cylinder is modeled by introducing complex wave numbers with normalized absorption coefficients independent of frequency, which holds for linear viscoelasticity [8, 46]. This behavior is typically encountered for various polymeric materials [4]. The normalized absorption coefficient for the longitudinal wave is given as [12] $\gamma_{L,PMMA} = 0.0119$, and the one for the shear wave is varied in the range $0 \leq \gamma_{S,PMMA} \leq 1$. Eqs.(1)-(4) are evaluated numerically by developing a MATLAB program used in the limit $0 < ka \leq 1$, where the anomalous effects have been observed. Adequate truncation of the series in Eqs.(1)-(4) has been applied such that $n_{max}$ largely exceeds

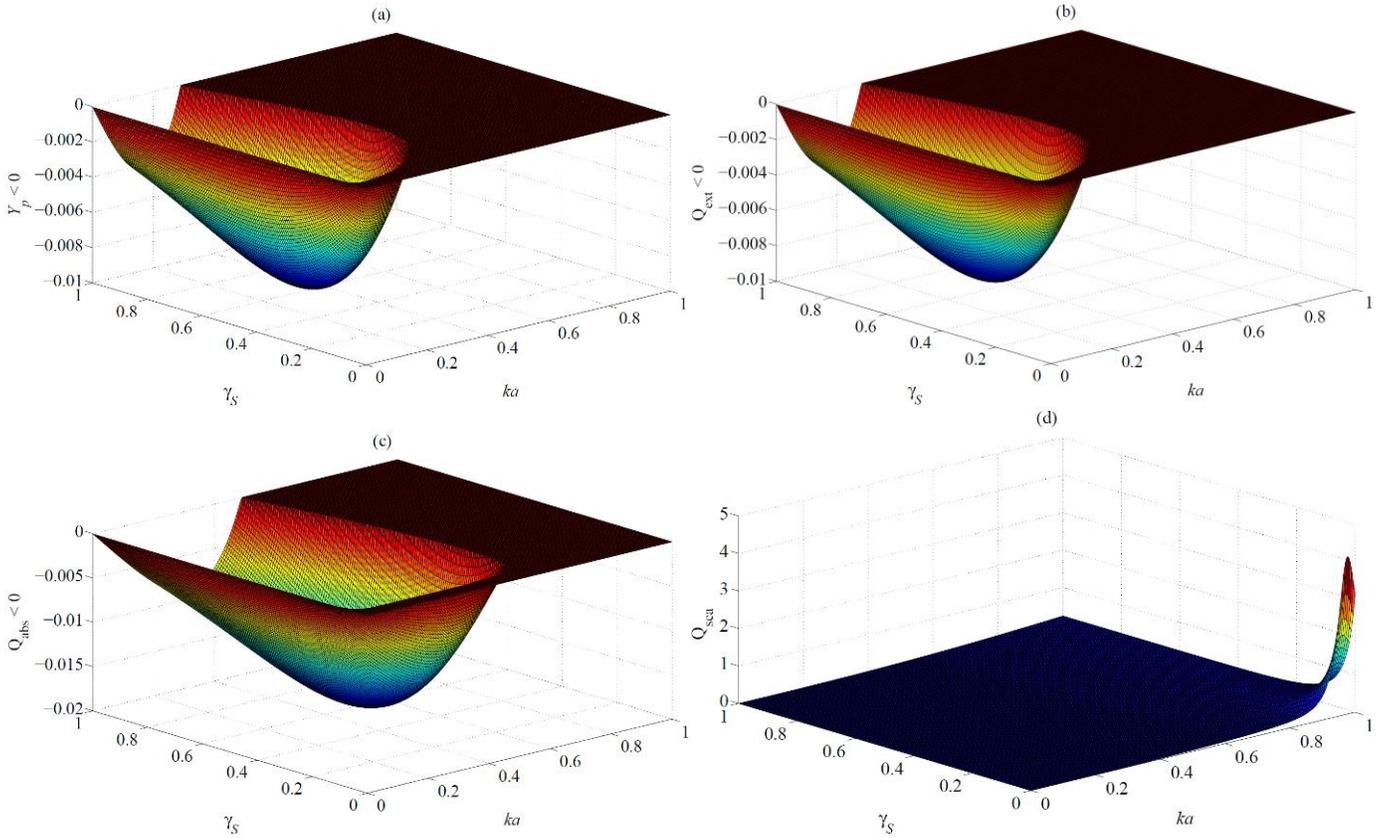

Fig. 2. The same as in Fig. 1 but for a VE-polyethylene (PE) cylinder.

$ka$ so that the numerical convergence of the series is warranted.

Panel (a) of Fig. 1 shows the computational plot of Eq.(1) in the ($ka$, $\gamma_{S,\text{PMMA}}$) ranges defined previously, over which the radiation force function $Y_p$ is found to be negative. The upper bound for $Y_p$ is set to zero for enhanced visualization of the negative force that occurs for $ka < 0.25$ and $0.05 < \gamma_{S,\text{PMMA}} \leq 1$. Note also that $Y_p$ has been computed for $\gamma_S > 1$ and has been found negative, however, these relatively large values for the shear-wave absorption coefficients may not be existent for this range of $ka$. It is important to note here that previous calculations of $Y_p$ for viscoelastic cylinders [16] did not display negative values because the normalized absorption coefficient for the shear-wave $\gamma_{S,\text{PMMA}}$ chosen in the computations for the selected materials did not exceed the minimal "critical" value required so as to produce a negative force.

Panel (b) of Fig. 1 shows the plot for the extinction efficiency given by Eq.(2) which also displays negative values in the same ($ka$, $\gamma_{S,\text{PMMA}}$) ranges noted for panel (a). This effect is expected since there is a direct connection of the force with the extinction (that is the sum of the scattering and absorption) cross-section [47-49].

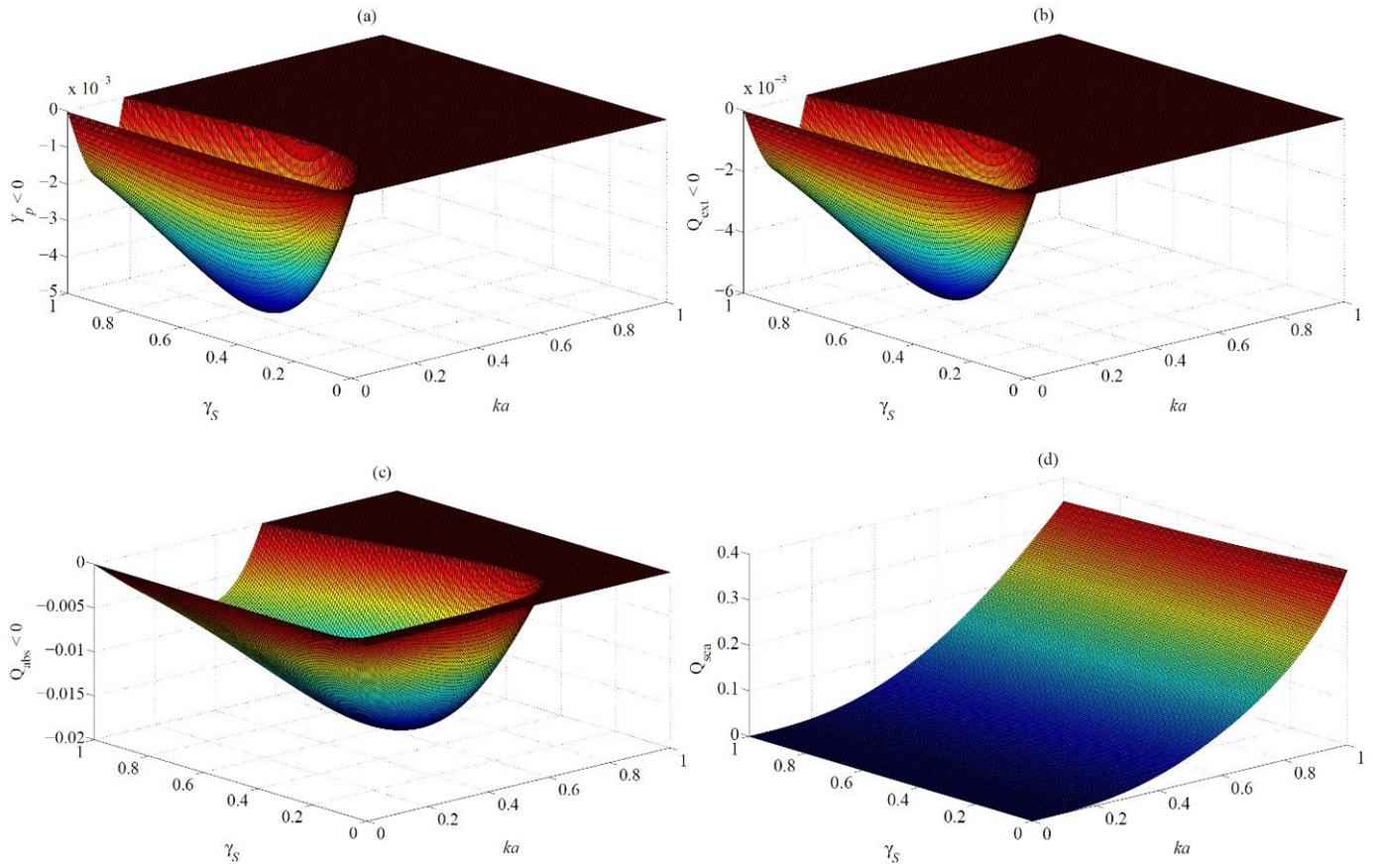

Fig. 3. The same as in Fig. 1 but for a VE-phenolic polymer (PP) cylinder.

Additionally, panel (c) of Fig. 1 displays the absorption efficiency given by Eq.(3), which is found to be negative. This indicates that the VE polymer cylinder acts as a steady-state source of energy (such as a perpetual motion system [50]), which is not permissible under the boundary conditions (i.e. continuity of stress and normal velocity at the boundary [51]) for a viscoelastic cylinder in a non-viscous fluid. Therefore, this result indicates that the values of $\gamma_{S,\text{PMMA}}$ for which $Q_{\text{abs}} < 0$ should be forbidden since the law of the conservation of energy $Q_{\text{ext}} = Q_{\text{abs}} + Q_{\text{sca}}$, is violated.

Panel (d) shows the scattering efficiency which is found to be positive for all the values of $\gamma_{S,\text{PMMA}}$.

The effects of changing the VE materials are shown in Figs. 2 and 3 for a PE ($\rho_{\text{PE}}$ = 957 kg/m$^3$, $c_{L,\text{PP}}$ = 2430 m/s, $c_{S,\text{PE}}$ = 950 m/s, $\gamma_{L,\text{PE}}$ = 0.0073) and a PP ($\rho_{\text{PP}}$ = 1220 kg/m$^3$, $c_{L,\text{PP}}$ = 2840 m/s, $c_{S,\text{PP}}$ = 1320 m/s, $\gamma_{L,\text{PP}}$ = 0.0119) cylinder, respectively [12]. From panels (a)-(c) of these figures, negative radiation force, extinction and absorption efficiencies have been found, similar to the panels of Fig. 1. This indicates that changing the material properties of the VE polymer cylinder causes minimal effects on the generation of these anomalies.

An attempt to further investigate the origin of these discrepancies consists of rewriting Eqs.(2)-(4) in terms of a scattering function $S_n = 2C_n + 1$, such that

$$Q_{ext} = \frac{1}{2ka} \sum_{n=0}^{\infty} \varepsilon_n \left(2 - S_n - S_n^*\right), \quad (5)$$

$$Q_{abs} = \frac{1}{2ka} \sum_{n=0}^{\infty} \varepsilon_n \left(1 - |S_n|^2\right), \quad (6)$$

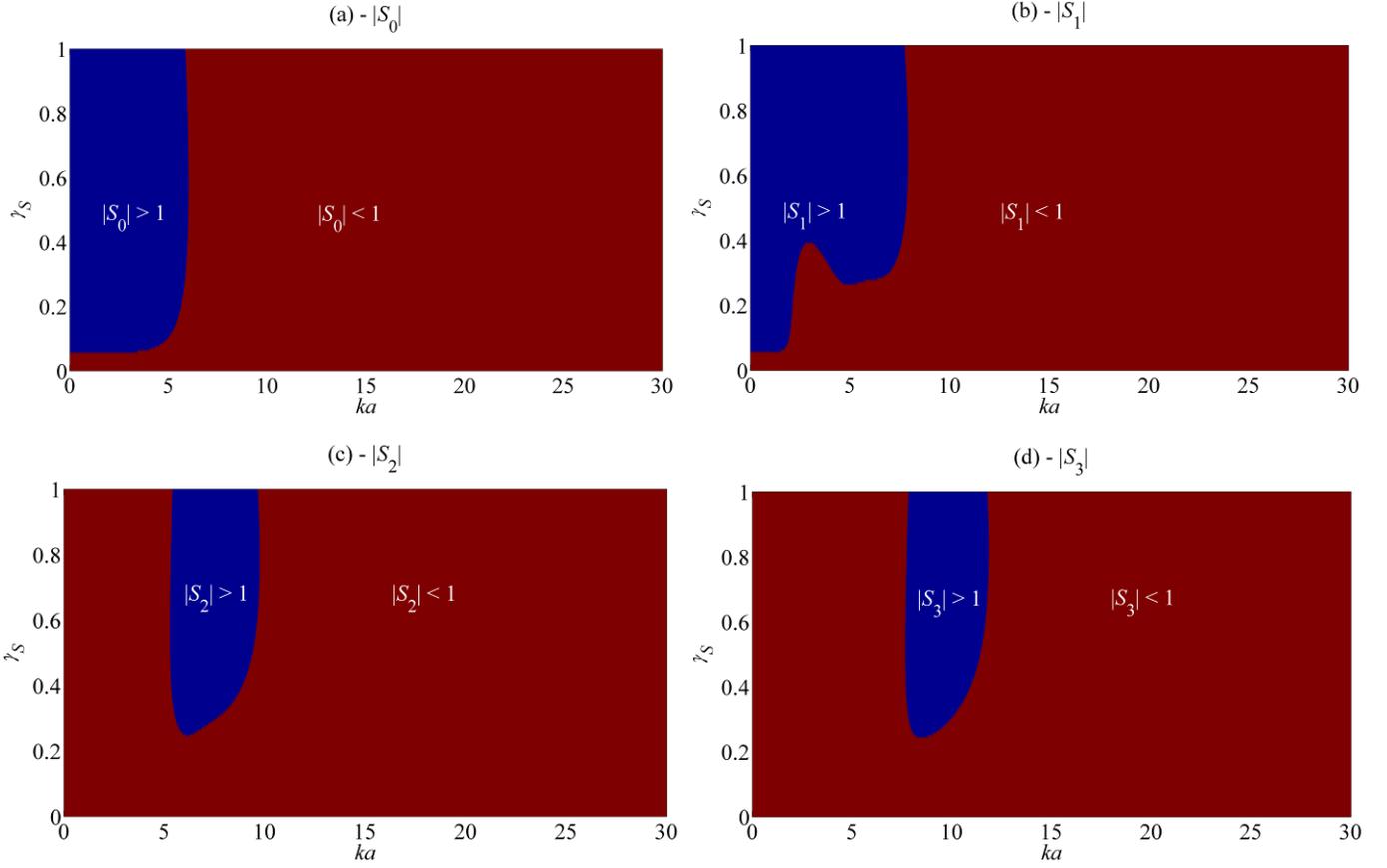

Fig. 4. The plots for the magnitude of the scattering form function for the monopole, dipole, quadrupole and hexapole partial-waves, shown in panels (a)-(d) respectively. The allowable physical values for $\gamma_S$ occur for $|S_n| \leq 1$.

$$Q_{sca} = \frac{1}{2ka} \sum_{n=0}^{\infty} \varepsilon_n |S_n - 1|^2, \qquad (7)$$

where the complex scattering function $S_n$ was initially introduced in the context of the nuclear scattering theory in particle physics [52, 53], and the superscript * denotes the conjugate of a complex number. In the acoustical context, the scattering function $S_n$ has some intrinsic properties such that $|S_n| = 1$ for elastic materials (i.e. $S_n$ is unitary), and $|S_n| < 1$ for VE materials [42]. As noticed from Eq.(6), the absorption efficiency cannot be negative, suggesting that the permissible values for the VE parameters should always be chosen such that $|S_n| \leq 1$.

Following this reasoning, the modulus of the scattering function $|S_n|$ is computed for the monopole ($n = 0$), dipole ($n = 1$), quadrupole ($n = 2$) and hexapole ($n = 3$) partial-waves, for a PP cylinder in water in the ranges $0 < \gamma_{S,PP} \leq 1$ and $0 < ka \leq 30$ at a fixed value of $\gamma_{L,PP} = 0.0119$. The results are displayed in panels (a)-(d) of Fig. 4, respectively. They show that the forbidden values for a VE-PP cylinder having $|S_n| > 1$ occur in different regions of ($ka$, $\gamma_S$). Comparison of panel (c) of Fig. 3 with those of Fig. 4 shows that the monopole ($n = 0$) term (corresponding to a "breathing" radially-symmetric vibrational mode) has the largest contribution, since the anomalies aroused at low $ka$ values < 1.

To further quantify the physical effect resulting from the monopole partial-wave term, the scattering coefficient $C_0$ for a viscoelastic cylinder in a non-viscous fluid (pp. 194-195 in [42]) is considered. Its expression is given by,

$$C_0 = -\frac{(\rho_0/\rho) x_L x_S^2 J_0(x) J_0'(x_L) - x J_0'(x)\left[2 x_L J_0'(x_L) + x_S^2 J_0(x_L)\right]}{(\rho_0/\rho) x_L x_S^2 H_0^{(1)}(x) J_0'(x_L) - x H_0^{(1)'}(x)\left[2 x_L J_0'(x_L) + x_S^2 J_0(x_L)\right]}, \quad (8)$$

where $\rho_0$ is the mass density of the surrounding fluid and $x = ka$, $x_L = (\omega/c_L)a(1+i\gamma_L)$, and $x_S = (\omega/c_S)a(1+i\gamma_S)$.

In the long wavelength limit $x = ka \ll 1$, it is possible to obtain an approximate expression of Eq.(8). This simplification provides some information in underlying the anomalous phenomenon observed here. In the long-wavelength limit, the following approximations for the cylindrical functions are used as [54],

$$J_0(x) \xrightarrow[x \ll 1]{} 1, \quad J_0'(x) \xrightarrow[x \ll 1]{} -(x/2), \quad H_0^{(1)}(x) \xrightarrow[x \ll 1]{} \left[1 + (2i/\pi)\ln(x)\right], \text{ and}$$

$$H_0^{(1)'}(x) \xrightarrow[x \ll 1]{} \left[-(x/2) + 2i/(\pi x)\right].$$

After some algebraic manipulation, the expression for $C_0$ is derived for a viscoelastic cylinder such that,

$$C_0\big|_{x \ll 1} = -\frac{(\rho_0/\rho) x_S^2 + x^2\left(1 - x_S^2/x_L^2\right)}{(\rho_0/\rho) x_S^2 \left[1 + 2i\ln(x)\right] + \left(x^2 - 4i/\pi\right)\left(1 - x_S^2/x_L^2\right)}. \quad (9)$$

For the case of an elastic (non-absorptive) cylinder, $x_L$ and $x_S$ become real numbers. Thus, Eq.(9) reduces to Eq.(16a) in [55] after neglecting the term $2i\ln(x)$, and the expression for the monopole coefficient in the long-wavelength limit shows a direct dependence on the 2D bulk modulus $B = \lambda_e + \mu_e$ [55], where the two elastic Lamé moduli of the cylinder are $\lambda_e = \rho(c_L^2 - 2c_S^2)$ and $\mu_e = \rho c_S^2$ (in units of Pa). However, for a viscoelastic cylinder, $x_L$ and $x_S$ are complex numbers. From Eq.(9), one notices the direct dependence of $C_0$ on the shear coefficient $x_S$, which in the case of the present numerical computations considered here, is about 2 to 6 times larger (in magnitude) than the compressional coefficient $x_L$. Therefore, the attenuation of shear waves inside the core material of the polymer (viscoelastic) cylinder strongly affects the monopole partial-wave ($n = 0$) mode, responsible for the generation of the anomalous unphysical effect of negative absorption. Unlike the case of a viscoelastic cylinder immersed in a non-viscous host fluid, the shear wave contribution to the monopole breathing mode has been previously observed in a different context, dealing with the scattering from a fluid-filled (spherical) cavity in a viscoelastic medium [56, 57].

The corresponding scattering function $S_0$ can be obtained from Eq.(8) [or from Eq.(9) in the Rayleigh limit] such that $S_0 = 2C_0 + 1$. For a given set of parameters for the VE material ($\rho$, $c_L$, $c_S$, $\gamma_L$) and the surrounding fluid ($\rho_0$, $c_0$), the suitable solutions for the shear-wave absorption coefficient $\gamma_S$ satisfying the conservation of energy law in accordance with the optical theorem should satisfy the equation,

$$|S_0| < 1. \qquad (10)$$

Eq.(10) is valid at all $ka$ ranges, and panel (a) of Fig. 4 shows the forbidden values (i.e. $|S_0| > 1$) for certain values of $\gamma_S$ at which the VE model fails to satisfy the conservation of energy law.

In conclusion, the present analysis suggests that the domain of validity for the linear VE model assuming a hysteresis-type of absorption in polymers in a non-viscous fluid hinges on the fact that the permissible values for the VE parameters should always be chosen such that the scattering function $|S_n| \leq 1$, which can be solved numerically. To further derive the appropriate constraints leading to adequate computations, consider the expression for the absorbed power $W_{abs}$ (given by Eq.(5) in [36]) and impose that it be positive (satisfying the condition of passivity [38]) as

$$W_{abs} = \left(-\int_S \langle p\mathbf{v}\rangle \cdot d\mathbf{S}\right) \geq 0, \qquad (11)$$

where $p$ and $\mathbf{v}$ are the total (i.e., incident + scattered) pressure and vector velocity fields, the symbol $\langle\ \rangle$ denotes time-averaging, $d\mathbf{S}$ (= $\mathbf{n}dS$) is the differential surface vector element, and $\mathbf{n}$ is the outward normal vector to the surface $S$. Let $\bar{\bar{\sigma}}$ denote the symmetric stress tensor, then $\bar{\bar{\sigma}}\mathbf{n} = -p\mathbf{n}$ as a consequence of traction continuity across the surface $S$. Thus, Eq.(11) becomes,

$$W_{abs} = \frac{1}{2}\mathrm{Re}\left\{\int_S \left(\bar{\bar{\sigma}} v_n^*\right) dS\right\} \geq 0, \qquad (12)$$

where $v_n = (\mathbf{v}\cdot\mathbf{n})$ is the normal velocity component.

Assuming a Kelvin-Voigt model of first order (i.e. linear), the stress tensor $\bar{\bar{\sigma}}$ for a solid VE isotropic material can be expressed as (Eq.(2.25) in [58])

$$\bar{\bar{\sigma}} = \left[\left(\lambda_e + \lambda_v \partial_t\right) \mathrm{tr}\bar{\bar{\varepsilon}}\right]\bar{\bar{\mathbf{I}}} + 2\left(\mu_e + \mu_v \partial_t\right)\bar{\bar{\varepsilon}}, \qquad (13)$$

where "tr" is the trace of a tensor, $\lambda_v$ and $\mu_v$ are the generalized viscosities (known also as the first and second viscous Lamé parameters, respectively, in units of Pa.s), $\partial_t = \partial/\partial t$, $\bar{\bar{\mathbf{I}}}$ is the second-rank unit tensor, and $\bar{\bar{\varepsilon}} = \frac{1}{2}\left(\overline{\overline{\nabla\mathbf{u}}} + \overline{\overline{\nabla\mathbf{u}^T}}\right)$ is the strain tensor, $\overline{\overline{\nabla\mathbf{u}}}$ is the displacement gradient tensor, and the superscript $T$ denotes the transpose. Provided that the deviatoric strain in 2D is $\overline{\overline{\varepsilon}}_d^{2D} = \overline{\overline{\varepsilon}} - \overline{\overline{\mathbf{I}}}\tfrac{1}{2}\mathrm{tr}\overline{\overline{\varepsilon}},$ the substitution of Eq.(13) into Eq.(12) leads to the following condition,

$$W_{abs} = \frac{1}{2}\int_S \mathrm{Re}\left\{\left[(\lambda+\mu)\overline{\overline{\mathbf{I}}}\,\mathrm{tr}\overline{\overline{\varepsilon}} + 2\mu\overline{\overline{\varepsilon}_d^{2D}}\right]v_n^*\right\}dS \geq 0, \quad (14)$$

where $\lambda = \lambda_e + \lambda_v \partial_t = \lambda_e - i\omega\lambda_v$, and $\mu = \mu_e + \mu_v \partial_t = \mu_e - i\omega\mu_v$, are the complex-valued Lamé moduli. Simplifying Eq.(14) further, the inequality is satisfied if,

$$\begin{cases}(\lambda_v + \mu_v) \geq 0 \\ \mu_v \geq 0\end{cases}. \quad (15)$$

For the VE model with hysteresis-type absorption, the coefficients of viscosities are, respectively,

$$\begin{cases}\lambda_v = \rho\left(\dfrac{c_L^2\left[1 - 1/(1+i\gamma_L)^2\right] - 2c_S^2\left[1 - 1/(1+i\gamma_S)^2\right]}{i\omega}\right) \\ \mu_v = \rho\left(\dfrac{c_S^2\left[1 - 1/(1+i\gamma_S)^2\right]}{i\omega}\right)\end{cases}. \quad (16)$$

Thus, Eq.(15) leads to the following constraints $\gamma_L \geq 0$ and $\gamma^{2D} \geq 0$, where,

$$\gamma^{2D} \approx \frac{c_L^2}{c_S^2}\frac{2\gamma_L}{\left(1+\gamma_L^2\right)^2} - \frac{2\gamma_S}{\left(1+\gamma_S^2\right)^2}. \quad (17)$$

For small values of $\gamma_L$ and $\gamma_S$, Eq.(17) can be replaced by,

$$\left.\left|\frac{\gamma_L}{\gamma_S}\right|\right|_{2D} \geq \frac{c_S^2}{c_L^2}. \tag{18}$$

Eq.(18) imposes a restriction on the values of the absorption coefficients for the VE cylinder (i.e. the 2D case).

For the 3D case, the deviatoric strain is $\overline{\overline{\varepsilon_d^{3D}}} = \overline{\overline{\varepsilon}} - \overline{\overline{\mathbf{I}}}\frac{1}{3}\mathrm{tr}\overline{\overline{\varepsilon}},$ leading to the constraints,

$$\begin{cases}(\lambda_v + \tfrac{2}{3}\mu_v) \geq 0 \\ \mu_v \geq 0\end{cases}. \tag{19}$$

Thus, Eq.(19) leads to the following constraints $\gamma_L \geq 0$ and $\gamma^{3D} \geq 0$, where,

$$\gamma^{3D} \approx \frac{c_L^2}{c_S^2}\frac{2\gamma_L}{\left(1+\gamma_L^2\right)^2} - \frac{8}{3}\frac{\gamma_S}{\left(1+\gamma_S^2\right)^2}. \tag{20}$$

For small values of $\gamma_L$ and $\gamma_S$, Eq.(20) can be replaced by,

$$\left.\left|\frac{\gamma_L}{\gamma_S}\right|\right|_{3D} \geq \frac{4c_S^2}{3c_L^2}. \tag{21}$$

Similar to Eq.(18), Eq.(21) imposes a restriction on the values of the absorption coefficients for a VE sphere (i.e. the 3D case).

Some empirical/experimental data for a high-density PE material which seem to violate the passivity conditions [i.e. Eqs.(18) and (21)] have been reported in Fig. 4 of [33]; for example, a PP cylindrical material with relatively weak absorption coefficients such that $\gamma_L = 0.001$ and $\gamma_S = 0.022$ does neither satisfy Eq.(18) nor Eq.(21). This suggests that such data appear to be inconsistent in some way. Note, however, that the other data sets reported in [33] have been also considered for samples of low-density PE and Lexan Plexiglas, which on the other hand satisfy the conditions given by Eqs.(18) and (21). An independent analysis [34] has also confirmed the presence of such discrepancies in Fig. 4 of [33].

It is also important to emphasize here that the normalized absorption coefficients obtained previously for a VE-PE material [4] (satisfying the passivity conditions) have adequately modeled the effect of acoustic attenuation inside an absorptive *sphere* or liquid-filled *spherical and cylindrical shell* in water from the standpoint of acoustic scattering and radiation force theories. Concerning the *coated/layered* sphere/shell model in the long-wavelength limit [35], the emergence of a negative force in plane waves is unphysical since the formulation used an incorrect coefficient for $\Lambda_{23}$ that is present in the matrix elements [25]. Its correct expression is given as [59] $\Lambda_{23} = \left(\frac{\rho_2}{\rho_3}\right)\left[\frac{c_{22}}{c_{32}(1+i\gamma_{22})}\right]^2$.

## 3. Conclusion

The results of the present analysis lead to the conclusion that any VE model describing the acoustic attenuation of incident acoustical arbitrary wavefronts (including plane waves and other beams) inside a polymer-type cylinder (or sphere), or multilayered objects in a non-viscous fluid must be validated and verified based upon the conservation of energy law using the (generalized) formalisms of the optical theorem for acoustical waves in cylindrical [36] or spherical coordinates [37]. Mathematical constraints for the viscoelastic moduli [given by Eqs. (18) and (21)] for a cylinder and sphere, respectively, are established stemming from the law of the conservation of energy, which must be enforced in order to obtain accurate results.

**Acknowledgement**

The first author gratefully acknowledges Prof. A. N. Norris (Dept. of Mechanical and Aerospace Engineering, Rutgers University, NJ, USA) for providing a comprehensive analysis on the absorption constraints on viscoelastic moduli [31], as well as Drs. J.P. Leao-Neto and G.T. Silva (Instituto de Fisica, Universidade Federal de Alagoas, Maceio, AL, Brasil) for pointing out the correction related to the coefficient $\Lambda_{23}$ for the layered sphere case.